# Actual Physics, Observation, and Quantum Theory

Tim Maudlin

Department of Philosophy, NYU

John Bell Institute for the Foundations of Physics

Abstract: Since its inception, quantum theory has been the subject of fierce interpretive controversy, which persists to this day. Disputed topics include the basic ontology and dynamics of the theory, the role (if any) of measurement, the meaning of probability, and the issue of non-locality. But there is yet another problem that has been largely ignored: how the theory makes contact with observational data. The problem is endemic to physics, and was discussed by Einstein in several places. In this essay, I discuss Einstein's general approach, how it applies to some quantum-mechanical phenomena, and why a central aspect of the solution might lead to novel and important new predictions.

Since the proposal of the "new quantum theory" by Heisenberg and Schrödinger, rigorous discussions its postulates have been difficult to achieve. Consider a comment made by Einstein in a letter to Schrödinger in 1928:

> The Heisenberg–Bohr tranquilizing philosophy—or religion—is so delicately contrived that, for the time being, it provides a gentle pillow for the true believer from which he cannot very easily be aroused.[1]

Similar expressions of Einstein's frustration with his colleagues can be found throughout his correspondence.

Every fundamentally new proposal for physics provokes some befuddlement and arguing at cross-purposes. But that discussions of quantum theory still resemble the situation after the destruction of the Tower of Babel is unprecedented in the history of science. Even today, it is hard to get agreement about most basic points and the differences between different approaches. This essay aims to exposit one neglected source of these difficulties.

Heisenberg relates an illuminating anecdote about a lecture he gave on matrix mechanics at the University of Berlin:

> In the spring of 1926, I was invited to address this distinguished body on the new quantum mechanics, and since this was my first chance to meet so many famous men, I took good care to give a clear account of the concepts and mathematical foundations of what was then a most unconventional theory. I apparently managed to arouse Einstein's interest for he invited me to walk home with him so that we might discuss the new ideas at greater length.
>
> On the way, he asked about my studies and previous research. As soon as we were indoors, he opened the conversation with a question that bore on the philosophical background of my recent work.

---

[1] Przibram, K. (1967) *Letters on Wave Mechanics: Schrödinger, Planck, Einstein, Lorentz* (New York: Philosophical Library), p. 31

"What you have told us sounds extremely strange. You assume the existence of electrons inside the atom, and you are probably quite right to do so. But you refuse to consider their orbits, even though we can observe electron tracks in a cloud chamber. I should very much like to hear more about your reasons for making such strange assumptions."

"We cannot observe electron orbits inside the atom," I must have replied, "but the radiation which an atom emits during discharges enables us to deduce the frequencies and corresponding amplitudes of its electrons. After all, even in the older physics wave numbers and amplitudes could be considered substitutes for electron orbits. Now, since a good theory must be based on directly observable magnitudes, I thought it more fitting to restrict myself to these, treating them, as it were, as representatives of the electron orbits."

"But you don't seriously believe," Einstein protested, "that none but observable magnitudes must go into a physical theory?" "Isn't that precisely what you have done with relativity?" I asked in some surprise. "After all, you did stress the fact that it is impermissible to speak of absolute time, simply because absolute time cannot be observed; that only clock readings, be it in the moving reference system or the system at rest, are relevant to the determination of time."

"Possibly I did use this kind of reasoning," Einstein admitted, "but it is nonsense all the same. Perhaps I could put it more diplomatically by saying that it may be heuristically useful to keep in mind what one has actually observed. But on principle, it is quite wrong to try founding a theory on observable magnitudes alone. In reality the very opposite happens. It is the theory which decides what we can observe.

> You must appreciate that observation is a very complicated process. The phenomenon under observation produces certain events in our measuring apparatus. As a result, further processes take place in the apparatus, which eventually and by complicated paths produce sense impressions and help us to fix the effects in our consciousness. Along this whole path—from the phenomenon to its fixation in our consciousness—we must be able to tell how nature functions, must know the natural laws at least in practical terms, before we can claim to have observed anything at all. Only theory, that is, knowledge of natural laws, enables us to deduce the underlying phenomena from our sense impressions.
>
> When we claim that we can observe something new, we ought really to be saying that, although we are about to formulate new natural laws that do not agree with the old ones, we nevertheless assume that the existing laws—covering the whole path from the phenomenon to our consciousness—function in such a way that we can rely upon them and hence speak of 'observations.'"[2]

The contemporary *verificationist theory of meaning* demanded a semantics akin to—but stricter than—Heisenberg's restriction of physical theories to statements in terms of observable quantities. Verificationism was roughly the doctrine that the *meaning* of a sentence is nothing but its means of verification or falsification, articulated in terms of an observation language. That language was to consist of unproblematically empirically verifiable or falsifiable "observation sentences". Heisenberg presents his approach not so much as a matter of semantic necessity but of good scientific practice, inspired by Einstein's own approach to space and time. It is often said that in Relativity Einstein replaced Newton's "metaphysical" account of space and time with empirical and operational definitions. If so, then talk of space-time structure would automatically become directly analyzable into the "observation language", just as Heisenberg desired.

But Einstein explicitly rejected this approach, as evidenced by Heisenberg's recollection.

---

[2] Heisenberg, W. (1972) *Physics and Beyond* (New York: Harper & Row), pp. 62-4

Any empirical science must be capable of some confirmation or disconfirmation via observation reports. But it is a long way from that triviality to the assertion that the content of *every* claim made in an empirical science must be translatable into an "observation language".

What exactly did Einstein mean when he said that *the theory* decides what we can observe?

One naïve thought immediately arises when considering that question. Suppose one is presented with a theory. There is no presumption that the theory is couched in terms of "observable quantities": perhaps it postulates unobservable entities such as atoms. The theory specifies an ontology (what exists) and a dynamics: an account of how the ontology behaves given by equations of motion. Such a theory has models: mathematical structures that are solutions of the equations of motion, and which represent ways the physical world could be (according to the theory). Programmatically, these solutions of the fundamental equations of motion play a similar role to "possible worlds" in the philosopher's possible world semantics. But if the fundamental terminology of the theory is *not* reducible to some "observation language", *how does the theory make any empirical predictions at all*?

Einstein had much to say about this problem, although his remarks are scattered through his writings. The remainder of this essay considers Einstein's account of the logical structure of empirical science, applying his approach to some current controversies. I contend that Einstein's suggestions provide the most reasonable path forward today.

We consider three Einsteinian texts. Each is well-known, and taken as a group they provide a coherent and somewhat surprising approach to the problem of extracting empirical consequences from a fundamental physical theory. None is explicitly about quantum theory, but the general outlook can be applied to it and helps when confronting serious *empirical* difficulties for "standard" quantum theory.

The first text is a comment made by Einstein in his 1948 essay "Quantum Mechanics and Reality":

If one asks what, irrespective of quantum mechanics, is characteristic of the world of ideas of physics, one is first of all struck by the following: the concepts of physics relate to a real outside world, that is, ideas are established relating to things such as bodies, fields, etc., which claim "real existence" that is independent of the perceiving subject—ideas which, on the other hand, have been brought into as secure a relationship as possible with the sense data. It is further characteristic of these physical objects that they are thought of as arranged in a space-time continuum. An essential aspect of this arrangement of things in physics is that they lay claim, at a certain time, to an existence independent of one another, provided these objects "are situated in different parts of space". Unless one makes this kind of assumption about the independence of the existence (the 'being-thus') of objects which are far apart from one another in space—which stems in the first place from everyday thinking—physical thinking in the familiar sense would not be possible. It is also hard to see any way of formulating and testing the laws of physics unless one makes a clear distinction of this kind. This principle has been carried to extremes in the field theory by localizing the elementary objects on which it is based and which exist independently of each other, as well as the elementary laws which have been postulated for it, in the infinitely small (four-dimensional) elements of space.

The following idea characterizes the relative independence of objects far apart in space (A and B): external influence on A has no direct influence on B; this is known as the 'principle of contiguity', which is used consistently only in the field theory. If this axiom were to be completely abolished, the idea of the existence of (quasi-) enclosed systems, and thereby the postulation of laws which can be checked empirically in the accepted sense, would become impossible.[3]

---

[3] Einstein, A. (1948) "Quantum Mechanics and Reality", *Dialectica* 2, Issue 3-4, pp. 320-321.

The main focus of the passage is a particular locality condition, which is also presupposed in the EPR argument of 1935[4] and is central to later results of John Bell. But the passage also contains an almost off-hand remark that deserves our consideration: 'ideas are established relating to things such as bodies, fields, etc., which claim "real existence" that is independent of the perceiving subject—ideas which, on the other hand, have been brought into as secure a relationship as possible with the sense data.' The crucial—and somewhat obscure—part of this comment is the last clause. Precisely here the theoretical account proposed by the physicist makes contact with experimental data and "observable quantities". Statements couched in the theoretical language have somehow been brought into "as secure a relationship as possible" with statements in the observation language. Unfortunately, Einstein gives us no clear account of exactly what that relationship is, or how it can be judged to be "secure", or how the degree of security might be improved.

Heisenberg wanted the security to be *absolute*. He demanded that the theory itself be constructed using only "observable quantities", which would make it easy to empirically test. Einstein firmly rejected that approach, but the alternative he had in mind is not readily apparent. All that is obvious is that Einstein did not seek—and thought one could not find—the connective tissue linking the theoretical description to the empirical data in pure logic or pure mathematics or semantic theory.

Einstein makes the same point in a letter he wrote to Maurice Solovine in 1952:

> As for the epistemological question, you completely misunderstood me; I probably expressed myself badly. I see the matter schematically in this way:

---
[4] Einstein, A., Podolosky, B., and Rosen, N. (1935) "Can Quantum-Mechanical Description of Physical Reality be Considered Complete?" *Phys. Rev.* 47 (10), 777-780

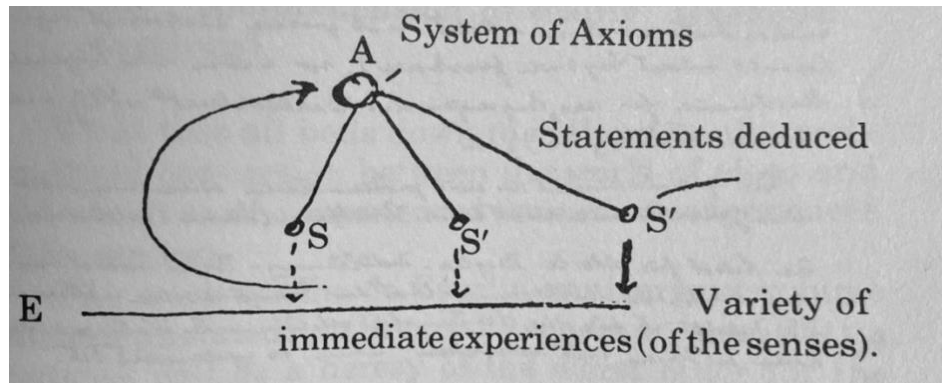

(1) The E's (immediate experiences) are our data.

(2) The axioms from which we draw our conclusions are indicated by A. Psychologically, the A's depend on the E's. But there is no logical route leading from the E's to the A's, but only an intuitive connection (psychological), which is always "re-turning".

(3) *Logically*, specific statements S, S', S'' are deduced from A; these statements can lay claim to exactness.

(4) The A's are connected to the E's (verification through experience). Closer examination shows that this procedure also belongs to the extralogical (intuitive) sphere, for the relation between the notions that show up S and the immediate experiences are not logical in nature.

But the relation between S's and E's is (pragmatically) much less uncertain[5] than the relation between the A's and the E's. (Take the notion "dog" and the corresponding immediate experiences.) If such a relationship could not be set up with a high degree of certainty (though it may be beyond the reach of logic), logical machinery would have no value in "the comprehension of reality" (example: theology).

---

[5] The translation in *Letters to Solovine* here has "certain" rather than "uncertain", but the original German handwritten text shows "unsicher" rather than "sicher".

> What this all boils down to is the eternally problematical connection between the world of ideas and that which can be experienced (immediate experiences of the senses).[6]

Some of Einstein's points are pellucid, reflecting what he told Heisenberg two and a half decades earlier. The axioms of the physical theory A are not couched in an "observation language" or solely in terms of "observable quantities". For example, a theory could ascribe a position and definite trajectory to the electrons in an atom even if neither of these could be "directly observed". From these axioms, by logic and mathematical reasoning, the consequences S, S', S'' etc. are derived, which—although equally not stated in any observation language—are somehow "intuitively closer" to observations or observational claims. That leaves us in need of further clarification about what this relation of "intuitive closeness" or "as secure a relationship as possible" is. We also face the question of what the character of the two languages—the language of A and the language of E—is.

We know from his remarks to Heisenberg that Einstein did not confine the quantities and entities mentioned in A to "observable quantities", and further did not think that the character of the entities postulated by A was derived by logic from the E's. The theoretical postulates are free inventions of the human mind—although they might be *inspired* by sense experience. It is now a truism that scientific theories cannot be *derived* from experience in any logical or algorithmic way, but involve the postulation of new sorts of "theoretical entities" such as electromagnetic fields or quantum states or "strings".

The S's, being logically derived from the A's, must also be in the theoretical language. This derivation makes use of the theoretical axioms, which is why Einstein insisted to Heisenberg that "it is the theory which decides what we can observe".

---

[6] Einstein, A. (1993) *Letters to Solovine* (New York: Citadel Press), pp. 137-139. Image used with permission of Philosophical Library.

So the gap to be bridged between the theory and the empirical evidence is the gap between the S's formulated in the theoretical language and the E's. The gap is bridged not by logic but by "intuition". What sort of concepts and terms are used in formulating the E's?

Einstein implies that the E's are stated *in terms of the "immediate experiences of the senses"*. But then the gap between the S's and the E's would yawn gapingly wide: as wide as the mind/body problem itself, with all the complications of human anatomy and neurophysiology piled on top of it. For the S's are both stated entirely in the language of physics and further can only cover as much of the physical world as the logical and mathematical derivation from the A's permits. And for *no empirical prediction ever made in the history of physics has the derivation from the axioms reached the interior of the human skull.* Newton could derive (with familiar idealizations and approximations) how a cannonball and a musket ball, released together above the surface of the Earth with negligible non-gravitational forces, would hit the ground together. But to trace the physics beyond that—to the light reflecting off the cannonball, into the eye of a nearby human, to the retina and ensuing neural activity, into the visual cortex, etc.—was and remains completely beyond the capacity of actual human scientists.

All of this goes without saying, but still needs to be said. For it is only somehow in the physical activity in the brain that physics finally makes a connection to *immediate experiences of the senses*. Connecting the one to the other would require a solution to the mind/body problem—which we have never had—but even were those connecting principles known the precise internal physical state of a human brain has never been among the S's that have been logically derived from any actual physical theory. So if the gap to be bridged in getting from the physical theory to the empirical data were there, there is no realistic prospect of ever even getting to the part of the duality that lies on the *physical* side of the mind/body gap.

Here the "Actual" in the title of this essay becomes relevant. The history of science is replete with discussions of some idealized kind of science, a science that humans never have and never will achieve. And there is a significant danger of unwittingly taking comments about what such an idealized science might possibly accomplish and applying them to *actual* scientific practice, where they can never obtain.

The sort of idealization I have in mind is commonly attributed to Laplace under the moniker *Laplace's Demon*, but Laplace was parroting remarks made half a century earlier by Rudger Bošković. Reflecting on the mathematical structure of Newtonian mechanics, Bošković wrote:

> Now, if the law of forces were known, & the position, velocity & direction of all points at any given instant, it would be possible for a mind of this type to foresee all the necessary subsequent motions & states, & to predict all the phenomena that necessarily followed from them. It would be possible from a single arc described by any point in an interval of continuous time, no matter how small, which was sufficient for a mind to grasp, to determine the whole of the remainder of such a continuous curve, continued to infinity on either side.[7]

Bošković, followed by Laplace, takes the literary license to discuss what is purely a matter of well-posed mathematical problems in terms of "predict[ing] all the phenomena" from initial conditions using a mathematically formulated theory such as Newton's. But if by "phenomena" one means "subjective conscious experiences", then no such thing could possibly be done by pure mathematical derivation from Newton's postulates. That is uncontestable. Either we have to simply drop all mention of "phenomena" from the discussion, or else find some non-psychologized version of "phenomena" to take the place of any talk of conscious observation.

Laplace's later version provides a step in the right direction. Laplace has:

> We ought then to regard the present state of the universe as the effect of its anterior state and as the cause of the one which is to follow. Given for one instant an intelligence which could comprehend all the forces by which nature is animated

---

[7] Quoted and translated in Kožnjak, B. (2015) "Who let the demon out? Laplace and Bošković on determinism", *Studies in the History and Philosophy of Science* 51, pp 42-52

and the respective positions of the beings which compose it - an intelligence sufficiently vast to submit these data to analysis - it would embrace in the same formula both the movements of the largest bodies in the universe and those of the lightest atom; for it, nothing would be uncertain and the future, as the past, would be present to its eyes.[8]

Laplace again leans on the rhetorical trope of what the demonic intelligence could be certain of—given unlimited analytical power—although the real topic has nothing to do with epistemology and everything to do with determinism. The rhetorical excess is understandable, but unfortunately suggests that the mathematical point has some bearing on how *actual human scientists use theories to make predictions, which are then subjected to empirical test*. And one thing sure: no human scientist has ever done or ever will do what the demon is stipulated to be capable of! Let's review some of the impediments.

First, the demon needs to have *a complete microscopic physical description of the entire universe and of its change over some period as initial data*. No such information ever has been or will be available to any actual scientist. Indeed, there are logical objections to the notion that any subsystem of the universe could contain a complete description of the entire universe in all detail.

Second, even if some human had such a description in hand, the calculation would be impossible to carry out. Humans can't even handle the three-body gravitational problem among point-masses in all detail.

Third, the range of observability of actual scientists is limited by horizons. If there is a huge electromagnetic pulse headed toward the Earth right now that will obliterate all life in two days, we *cannot* know about it in time to predict it.

Fourth, the number of physical degrees of freedom in even a box of gas is so large, and the detailed state of the gas so delicate, that there will never be any means to determine it to the degree of precision needed, much less to do that without massively disturbing it (even apart from any considerations of quantum mechanics).

---

[8] Laplace, P. S. (1902/1814). A philosophical essay on probabilities (F. W. Truscott & F. L. Emory, Trans.). London: John Willey & Sons, p. 4

Fifth, the dynamical laws might not be deterministic. In such a case one would presumably only expect probabilistic predictions rather than predictions with certainty.

At a real, practical level, the five points above guarantee that there never has been and never will be an actual Boškovićian or Laplacian demon. Actual scientific practice just does not work that way. Still, one might argue, we can accept the demon as an *idealized* scientific investigator, which the human scientist can strive to emulate as far as possible. In service of this idea, consider some of the conceptual and calculational shortcuts, approximations, and ameliorations the real scientist might use.

Regarding the unlimited extent of the physical state of the entire universe, the scientist can try to find or construct isolated or quasi-isolated systems, or systems such that the physical influence of items outside it can be well-approximated somehow. Einstein regarded the existence of such systems as so essential for making scientific predictions that he posited a form of *causal and ontological locality* as a transcendental prerequisite for science to be possible. That is one of the reasons he regarded "spooky action-at-a-distance" with such revulsion. He worried that any theory postulating such action—which would make causally (approximately) isolated systems impossible—would undercut the entire scientific method. Quantum theory stands as a counter-example to that worry, although it is a delicate task to explicate exactly why the theory is so testable despite its non-locality.

Another ameliorating technique uses statistical methods. The precise physical state of a box of gas will never be knowable in practice. But statistical mechanics explains how—given the right sort of dynamics—probabilistic predictions about behavior of *statistical averages* can be made. The predictions are not made with certainty, but the uncertainties can sometimes be reduced to practical insignificance. These techniques do not require the predictor to calculate the *exact* phase trajectory of even one phase point—much less all of them—and can be robust against a wide variety of variations in the micro-dynamics. One foreswears the demon's ability to precisely predict details, but gains traction over the near-certain prediction of statistical quantities.

Another ameliorating maneuver is hinted at by Laplace. Although we naturally regard the demon as calculating the *precise future or past microstate of a system* (since the fundamental dynamics of the system are formulated in those terms), Laplace notes that the demon will predict "both the movements of the largest bodies in the universe and those of the lightest atom". From the demon's point of view, once the movements of all the atoms have been calculated the movements of the large bodies follow. In the order of ontological dependency the atoms come first and the agglomerations of them follow. But with respect to our problem of actual scientists deriving empirical predictions, the order is the reverse. To put it bluntly, claims about the motions of the large bodies can be part of the content of claims in the "observation language," while claims about the motion of a single atom cannot. Indeed, it is exactly here—at the level of the positions and motions of (relatively) large bodies—that the logical gap between the S's and the E's reduces to near nothing. From the axioms of Newtonian gravitational theory and some simple idealizations one can calculate that a uniform iron sphere the size of a musket ball and a uniform iron sphere the size of a cannonball will (neglecting air resistance) fall with the same acceleration on the surface of the Earth. That S-claim can be brought into a "secure relationship" with E-claims produced by observers.

Carrying out this reconciliation between theory and evidence requires adjustments and constraints on both sides of the gap. On the side of theory, the constraint seems trivial and "purely logical": instead of predicting the motions of individual atoms, the theorist must predict the motions of large ("macroscopic") collections of them. The constraint is indeed trivial for the demons but a godsend for real human scientists, who could not possibly carry out the detailed, exact microscopic calculations but—via various approximations and idealizations—can get their hands on predictions for macroscopic bodies. The required approximations and idealizations are worthy of close consideration.

On the side of the E's, the observational evidence, the adjustment is in one way simple but in another radical. Einstein's comment to Heisenberg about "further processes take place in the apparatus, which eventually and by complicated paths produce sense impressions and help us to fix the effects in our consciousness" and his talk to Solovene of "the eternally problematical connection between the world of ideas and that which can be experienced (immediate experiences of the senses)" both suggest that what he had in mind by the E's were *claims about subjective conscious experience*. If so, then contact between the S's and the E's as a realistic practical matter will be lost. If any scientist were even required to model the complete physics of the apparatus and brain of the observer then no predictions would ever be made.

The situation is different if the E's describe dispositions and motions of middle-sized objects such as cannonballs. That would make the S's side of the gap realistically accessible by mathematical derivation from the axioms via simplifications and idealizations. From the theoretical axioms one derives the trajectory of a cannonball-sized uniform solid sphere, which is then *directly* compared with observation reports. Not only would the gap between the S's and the E's become bridgeable, but the E's themselves become easily intersubjectively testable, unlike claims about subjective experiences. Descartes may know with great certainty how things subjectively appear to Descartes, but for everyone else it is much easier to establish whether the musket ball and cannonball hit the ground together than to determine any facts about how things seem *to Descartes.* Taking intersubjective testability as a central feature of empirical science, this choice for the observation sentences is vastly superior to descriptions of subjective conscious experiences.

Choosing middle-sized-observable-thing-language for the E's, then, solves conceptual problems and makes the theorist's task of deriving closely related S's tractable using simplifications, approximations, and idealizations. This seems like the most charitable way to interpret Bohr's dictum that even when employing the quantum-theoretic apparatus *one must describe the experimental situation and outcomes using "classical concepts"*:

> It lies in the nature of physical observation, nevertheless, that all experience must ultimately be expressed in terms of classical concepts, neglecting the quantum of action. It is, therefore, an inevitable consequence of the limited applicability of the classical concepts that the results attainable by any measurement of atomic quantities are subject to an inherent limitation.[9]

Bohr cannot mean by "in terms of classical concepts" that one uses classical *mechanics* to describe the behavior of the experimental apparatus: if that were so there would be no need for quantum theory at all! His intention is that the language in terms of which the "empirical" or "observable" phenomena are described must lend itself to repeated intersubjective testing, which requires being couched in terms of the positions, shapes, constitution, and motion of macroscopic objects. Bell praised this aspect of Bohr's philosophy, citing[10] Bohr's comment in 1949:

> …it is decisive to recognize that, *however far the phenomena transcend the scope of classical physical explanation, the account of all evidence must be expressed in classical terms*. The argument is simply that by the word "experiment" we refer to a situation where we can tell others what we have done and what we have learned and that, therefore, the account of the experimental arrangement and of the result of the observations must be expressed in unambiguous language with suitable application of the terminology of classical physics.[11]

Since classical physics—as Einstein insisted—was couched entirely in terms of what Bell called "local beables", it follows from Bohr's observation that *the description of experimental conditions must of necessity be given in terms of the arrangement and behavior of some local beables*. And if that arrangement and behavior is to be *observable*, then it must describe local beables at macroscopic scale. *That* is the language in which Einstein's E's must be couched.

---

[9] Bohr, N. (1929) "The Quantum of Action and the Description of Nature," originally published in German in *Naturwissenschaft*, vol 17, 1929. First published in English in 1934 by Cambridge University Press.
[10] Bell, J. (1976) "The Theory of Local Beables", *Epistemological Letters* (March edition). Reproduced as Chapter 7 of *Speakable and Unspeakable in Quantum Mechanics* (1987) Cambridge: Cambridge University Press.
[11] From "Discussion with Einstein on Epistemological Problems in Quantum Physics" in *Albert Einstein Philosopher-Physicist*, ed. By P. A. Schilpp, LaSalle:Open Court, Volume One, p. 209.

Bell's point in "The Theory of Local Beables" can now be stated succinctly: if the E's must be given in terms of macroscopic local beables, then the "intuitive" and "secure" connection requires that the S's—which are logically derivable from the fundamental axioms A—*must themselves contain reference to some local beables*. It is exactly there that the theoretical and observational descriptions make "intuitive" contact. Hence the A's themselves must either postulate some local beables or else somehow provide an ontology from which they can "emerge".

In "Against Measurement"[12], Bell pushes this line of thought further. Bohr insisted that a physical theory must use classical language at macroscopic scale in order to accommodate experimental reports. But at the same time, Bohr rejected the postulation of any local beables at *microscopic* scale, where "the quantum description" was supposed to be deployed. To Bell this position made no sense, not least because there is no principled sharp distinction between microscopic and macroscopic. But more than that, since Democritus physicists have pursued a simple idea: macroscopic (local) items *are large collections of microscopic local items*. This idea is so simple and compelling that it is hard to understand what alternative there might be: how could macroscopic local objects of the sort adverted to in the descriptions of experiments and their outcomes exist if not as large collections of microscopic local constituents? Even if the theoretical description contains things *in addition* to the microscopic local beables (such as quantum states), it is from the behavior of these microscopic local entities that the behavior of the macroscopic local objects follows automatically.

Bell puts it this way:

> It seems to me that the only hope of precision with the dual $(\Psi, x)$ kinematics is to omit completely the shifty split [between the "classical" and "quantum" realms], and let both Ψ and $x$ refer to the world as a whole. Then the $x$s [i.e. the local beables] must not be confined to some vague macroscopic scale, but must extend to all scales. In the picture of de Broglie and Bohm, every particle is attributed a position $x(t)$. Then instrument pointers—assemblies of particles—*have* positions, and experiments *have* results. The dynamics is given by the world Schrödinger equation plus precise 'guiding' equations prescribing how the $x(t)$s move under the influence of Ψ. Particles are *not* attributed angular momenta, energies, etc., but *only* positions as functions of time. Peculiar 'measurement' results for angular momenta, energies, and so on, emerge as pointer positions in appropriate experimental set-ups.[13]

---
[12] Reproduced as Chapter 23 of *ibid*.

Bell's approach, presented in terms of the programmatic role of local beables in the articulation of a physical theory, seems to solve completely the problem of connecting the S's to the E's. It accounts for how physical theories can be subjected to experimental test even though the postulates of the theory *cannot* be stated solely in terms of the observation language (and even if the theory postulates some non-local beables that *cannot* be described in terms of local beables at all). The microscopic local beables playing this programmatic role need not be particles. Local microscopic fields, for example, or point events ("flashes"), or strings, could do just as well, so long as how observable macroscopic objects are constituted out of the microscopic elements is clear.

Bell's methodological solution seems to provide the resources needed to solve Einstein's gap between the language of the theory and the language of the experimental reports without embracing Heisenberg's demand that the theory be presented solely in terms of "observable quantities", and avoids Einstein's hopeless suggestion that the evidence be presented in terms of subjective conscious experience.

But there is still a fly in the ointment.

---

[13] Ibid, p. 228.

Ironically, the fly I have in mind was articulated by Einstein but not in the context of discussing quantum theory. Rather, he noted a conceptual difficulty in his own presentation—in 1949—of the Theory of Relativity! Since Relativity is not generally regarded as having any deep unresolved conceptual issues, it is worthwhile to reflect carefully on the point Einstein was making. His discussion clarifies some of the issues articulated above.

The passage occurs after Einstein has presented an account of Relativity in terms of the behavior of clocks and rods and light rays in Minkowski space-time:

> First, a remark concerning the theory as it is characterized above. One is struck [by the fact] that the theory (except for the four-dimensional space) introduces two kinds of physical things, i.e., (1) measuring rods and clocks, (2) all other things, e.g., the electromagnetic field, the material point, etc. This, in a certain sense, is inconsistent; strictly speaking measuring rods and clocks would have to be represented as solutions of the basic equations (objects consisting of moving atomic configurations), not, as it were, as theoretically self-sufficient entities. However, the procedure justifies itself because it was clear from the very beginning that the postulates of the theory are not strong enough to deduce from them sufficiently complete equations for physical events sufficiently free from arbitrariness, in order to base upon such a foundation a theory of measuring rods and clocks. If one did not wish to forego a physical interpretation of the coordinates in general (something which, in itself, would be possible), it was better to permit such inconsistency — with the obligation, however, of eliminating it at a later stage of the theory. But one must not legalize the mentioned sin so far as to imagine that intervals are physical entities of a special type, intrinsically different from other physical variables ("reducing physics to geometry," etc.).[14]

Einstein's remark about the conceptual inconsistency of the presentation is clearly correct. The basic ontology employed in the presentation—the ontology deployed without being given any deeper ontological or physical analysis—contains *not merely* particles and fields *but also* items identified as "clocks" and "rods". But physicists know that clocks and rods are just complicated configurations of more basic physical constituents. So one can't simply stipulate how the clocks and rods behave without potentially running afoul of their further analysis into particles and fields.[15]

---
[14] *Albert Einstein Philosopher-Scientist*, op. cit., Volume 1, p. 59-61.

Einstein's point is that as a *practical* matter the analysis of clocks and rods as complex physical systems *could not* be carried through at the time: the requisite physical theory of matter did not exist. The only practical expedient allowing physics to progress was to *stipulate* what these devices do and use that stipulation to make testable predictions. But the stipulation should be made self-consciously, understanding that an obligation has been incurred (one that can only be discharged later, when the theory of matter has become sufficiently advanced). *There was no other way that physics could proceed*. Consider, in this case, exactly what these stipulations and their attendant obligations are.

Although Relativity is commonly presented in terms of the behavior of clocks and rods, it simplifies matters and makes the exposition more realistic by omitting the rods altogether. "Spatial distances" between events and objects are introduced via radar-ranging: a beam of light (in a vacuum) is emitted at one point, reflects off an object and returns, while an inertial clock measures the time $\Delta t$ from emission to return. A "distance" of $\frac{\Delta t}{2c}$ from the ranged object to the inertially moving emitter can be defined without mentioning rods.

For this observational procedure to make contact with the fundamental Relativistic structure of space-time, what must the clock do? It must have three features. First, its ticking must be proportional to the (theoretically defined) proper time along its trajectory. Second, the clock must start when the light is emitted. Third, it must stop—or mark the time—when the light returns. The clock is presumed to have these features when observationally testable claims about clocks in particular circumstances are "derived" in a Relativistic setting.

---

[15] To paraphrase Bell, if one makes postulates, rather than definitions and theorems, about the behavior of "measuring instruments" then one commits redundancy and risks inconsistency…..even regarding clocks and rods in Relativity.

It is obvious that compared to actual, observable clocks these are idealizations. No real clock is going to mark time in *exact* proportion to the proper time, and in any real clock there will be some lag between the starting of the clock and the emission of the light and the stopping of the clock and the reception of the light. But calibration of a real clock's departure from the ideal can be determined by both theoretical and experimental means.

*Theoretically* justifying the claim that a real clock (approximately) ticks in proportion to the proper time requires a detailed account of the clock mechanism. Physics texts often attempt this sort of analysis by reference to "light clocks", which work by having a beam of light bounce back and forth between two mirrors and counting the ticks when it hits one of them. This mechanism has the advantage of being somewhat analyzable in terms of fundamental physical principles, but the severe disadvantage of not resembling any actual clock ever used to take data. This illustrates the difficulty of getting a pure theoretical description to make contact with obtainable data in a conceptually rigorous way. Einstein, in the quote above, acknowledges the situation. But given the state of play at the time, all he could do was introduce fictitious ideal "clocks" into the theoretical description—with no description of how they worked but only of what they did—leaving for later a more realistic account.

*Given* this stipulated characterization of the clocks—also stipulated to be macroscopic—one then derives predictions observable phenomena from the theoretical description. Those predictions can be tested against real data from the lab. And given how holistic confirmation works, we can say that if the predictions turn out to be accurate that counts as some confirmation that *both* the basic theory *and* the stipulations are not too far off the mark. Conversely, if the predictions come out wrong then *either* the fundamental theory *or* the stipulations (or both) must be erroneous. Partisans of the theory have the option of blaming predictive failures on the stipulations: it is not the theory but the assumptions about the instrumentation that need improvement. *This is the only way that confirmation and disconfirmation proceed in actual*

*physics. Physicists have never been able to represent and analyze the entire experimental situation up to macroscopic scale using only the resources of the fundamental theory*. We might reasonably call this "the measurement problem as it appears in Relativity". Experiments that produce actual data always involve the use of an apparatus *that is not practically amenable to representation or calculation in terms of the fundamental theory*: the apparatus is *always* treated in some simplified, unphysical manner. This can be due (as for Einstein) to the absence of any existing theory of the fundamental physics of the apparatus, or due to the inaccessibility of a precise physical description of the apparatus, or even to the practical impossibility of solving equations that describe the apparatus at microscopic scale. Recall the moral of our discussion of Bošković's and Laplace's demons: although they are described picturesquely in terms of what they can predict or retrodict, the point Bošković and Laplace were making was about only the mathematical form of the fundamental dynamical equations. Actual scientists never have and never will make predictions following the method of the demons, not even approximately. We need to think clearly about what actual scientists are able to do, and how.

Einstein made a similar point about the practical necessity of unrealistic representations when discussing General Relativity. Concerning the General Relativistic Field Equation he wrote:

> It is sufficient—as far as we know—for the representation of the observed facts of celestial mechanics. But, it is similar to a building, one wing of which is made of fine marble (left part of the equation), but the other wing of which is built of low-grade wood (right side of the equation). The phenomenological representation of matter is, in fact, only a crude substitute for a representation which would correspond to all known properties of matter.[16]

Einstein again acknowledges that deriving testable predictions from the theory requires use of an unrealistic representation of matter, a representation rooted in idealization, appeals to classical physics, and mathematical tractability. The resulting mathematical models (solutions of the field equations) were known from the beginning not to provide perfectly accurate and faithful

---

representations of the real physical situation, but still might well be *good enough for the purposes at hand*. The only way to determine that was to come up with some solvable models and compare the predictions obtained from them with observational data. If the fit is close enough the procedure is justified as a promising ongoing research programme. One aspect of the research is to reduce—and ideally eliminate—the manifest crudity of the representation of matter. This procedure bears no resemblance to the operations of the fictitious demons.

What is of paramount importance, given the present state of physics, is that Einstein's observation about Relativity fits our current situation with respect to quantum-mechanical phenomena. There seems to be a widely shared—but radically mistaken—view that "quantum theory" as we presently have it cleanly makes extremely accurate predictions for all the experiments we can carry out in the lab using available technology. Or at least, any practical difficulties we may have in such situations stem from the limitations that prevent us from being demons: finite computational power and analytical abilities. The predictions, one might say, are *already there implicitly in the equations*, and we just need to make improvements to our ability to tease them out. But that is not the situation, even with respect to one of the most iconic and simplest experiments, namely the two-slit experiment.[17]

Surely, one thinks, of all the quantum-mechanical experiments one could mention the one over which we have the greatest theoretical and calculational control is the two-slit experiment! And further, we not only know how to make predictions of the observable phenomena, we have made experiments to test the predictions, and the predictions are correct. But to what degree is that accurate? Let's walk through some of the steps that go into making the calculation.

Suppose—to make the situation cleaner—we confine a single particle in a trap and cool it to its ground state, so we have control over its initial wavefunction. At a predetermined time we drop one side of the trap and let the wavefunction evolve toward a barrier with two slits.

---

[17] For an enlightening modern overview of this experiment, see Das, S., Deckert, D.-A., Kellers, L., Krekels, S. and Struyve, W. (2025) "Double slit experiment revisited", *Annals of Physics* 479, id 170054

Beyond the barrier, there is a screen. We want to predict, among other things, the distribution of spots on the screen after many repetitions. How do we analyze this experimental set-up theoretically?

At first, it seems simple. We have the initial wavefunction in the trap and the Schrödinger equation for its dynamics. The barrier is a complicated constellation of particles but we do not attempt to model them individually. Rather, that part of the experimental apparatus is represented by steep and high potential functions, chosen to be mathematically tractable. Given the boundary condition imposed, the Schrödinger evolution of the wavefunction can be solved and on the far side of the barrier the usual interference of the wavefunction occurs. That takes care of everything…except the screen. But without the screen somehow incorporated into the model no empirical predictions at all can be derived.

How can we theoretically handle the screen? There are two issues, each of which is important. The physical presence of the screen will, at some point, influence the wavefunction of the particle and change it from what it would have been without the screen (including entanglement with the screen). Even more important is the physical function of the screen *as a device where observable marks form*. Those marks constitute the outcome of the experiment, so accounting for the data just is accounting for the locations of the marks. How are we to do that?

The standard approach to "measurement" in quantum theory famously makes a mystery out of all this. The actual technique for making the predictions involves a series of steps: first, identifying the screen as a "measurement device" which "measures" some "observable quantity", such as "position" or "momentum". Second, the mathematical identification of the "observed quantity" with some Hermitian operator or POVM *on the Hilbert space of the "observed system".* Third, the use of that operator to extract a probability measure over "outcomes", e.g. by expressing the wavefunction of the "observed system" as a superposition of eigenstates of the operator and then squaring the amplitude of each "outcome".

Even the most casual consideration shows that the physical "measurement device" is not being modeled physically in this procedure: it has been completely abstracted away. Nothing in the method provides the slightest clue of how to account for the effect of the physical presence of the device on the wavefunction of the target system, and hence on the outcomes. The abstraction is so extreme that there is no way to incorporate that information into this sort of calculation: physically different systems that all count as "position measuring devices" or "momentum measuring devices" enter into the calculation in exactly the same way—via the Hermitian operator or POVM—so their precise physical characteristics *cannot* make any difference to the predictions. It is as if the "measuring devices" work by magic.

No effort is made to take into account the internal physical structure of the "measurement device". Just as with the "clocks" and "rods" in Einstein's presentation of Relativity, all of that physics has been elided. And just as with Einstein's presentation, the resulting story is conceptually incoherent, since the device—whatever it is—operates in virtue of its physical structure, and ought in principle to be amenable to some degree of physical analysis.

Since the "measurement device" has not been integrated into the physical description in a principled way, there is also no way to take into account the effect of the interaction on the target system after the interaction is over. Often, it is asserted that the post-interaction state of the target system will be the eigenstate of the operator corresponding to the "observed outcome", but for most real experimental situations that will be flatly false (e.g. when a photon has been completely absorbed by a screen). Where it might be approximately true, no account is given of how such a result would come about via the physics of the device.

Finally, this predictive technique depends critically on the association of the "measurement device" with some Hermitian operator or POVM, and no clue has been given about how to do that in a principled way. Given a *complete* quantum-mechanical characterization of the device, we are given no guidance about how to determine what—if anything!—it "measures" or which—if any—Hermitian operator or POVM ought to be associated with it. The association is typically made via some appeal to *classical* physics: in a classical setting, the experimental set-up would be regarded as a way to get information about the position or momentum of a particle, and therefore(?) we are entitled to regard it as a "position measurement" or a "momentum measurement" *even though all of classical physics including the existence of any particle is being rejected*. The logical justification of such a procedure is, to say the least, suspect.

How can we do better? Obviously, by treating the experimental apparatus as a physical entity and the interaction with the target system as just the same sort of physical interaction as any other. This is the principled way to avoid the "measurement problem": don't mention "measurement" anywhere in the fundamental principles of the theory *or in the practical application of the theory that is used to make predictions*. Such a treatment would avoid the major interpretational difficulties associated with "standard quantum theory".

Unfortunately, this observation all too easily suggests that the *only* way to avoid these problems is to go all the way to the demonic method: model the entire situation globally using the fundamental physics and then calculate—as best one can—what the predictions for the outcome should be. The difficulty is that the strategy cannot be implemented for *actual* physics done by *actual* scientists. The sorts of stipulations-cum-promissory-notes that Einstein adverts to *must* be used if actual numbers are to be forthcoming. The question is not whether such expedients will be used but rather how and where, what the acknowledged imperfections in the stipulations are, and how they might be reduced in the future.

Given an approach to quantum theory according to which the wavefunction is complete[18] one hardly knows where to begin. I will not even try to speculate. But in the case of theories with a richer fundamental ontology the conceptual path forward is reasonably clear, and worthy of being articulated and considered.

In the standard two-slit set-up, the standard theory is hamstrung because *it positively denies that there are any particles that arrive and interact with the screen at any particular time and place.* All that "arrives at the screen" is (as it were) the system's wavefunction, and that washes over the screen along its entire spatial extent and over a substantial period of time. How one is supposed to get from that to the formation of a spot on the screen at a particular place and time—which is what we think we can observe and collect data about—is a puzzle that holds the standard "measurement problem" at its core.

---

[18] By "complete" here, all I mean is that every physical characteristic of the system is reflected—somehow—in the wavefunction ascribed to it.

In a theory that postulates *actual particles* that follow *definite trajectories* and arrive near the screen at particular times and places, the situation is not nearly so desperate. But neither is it entirely straightforward. Using the standard pilot wave theory, with the initial wavefunction of the trapped particle given and the rest of the apparatus *sans* screen represented by a steep potential, it is possible to solve for the exact trajectory of a particle that starts at a given location. This calculation yields the familiar diagram of trajectories in the two-slit experiment according to the Pilot Wave theory (Figure 1).[19]

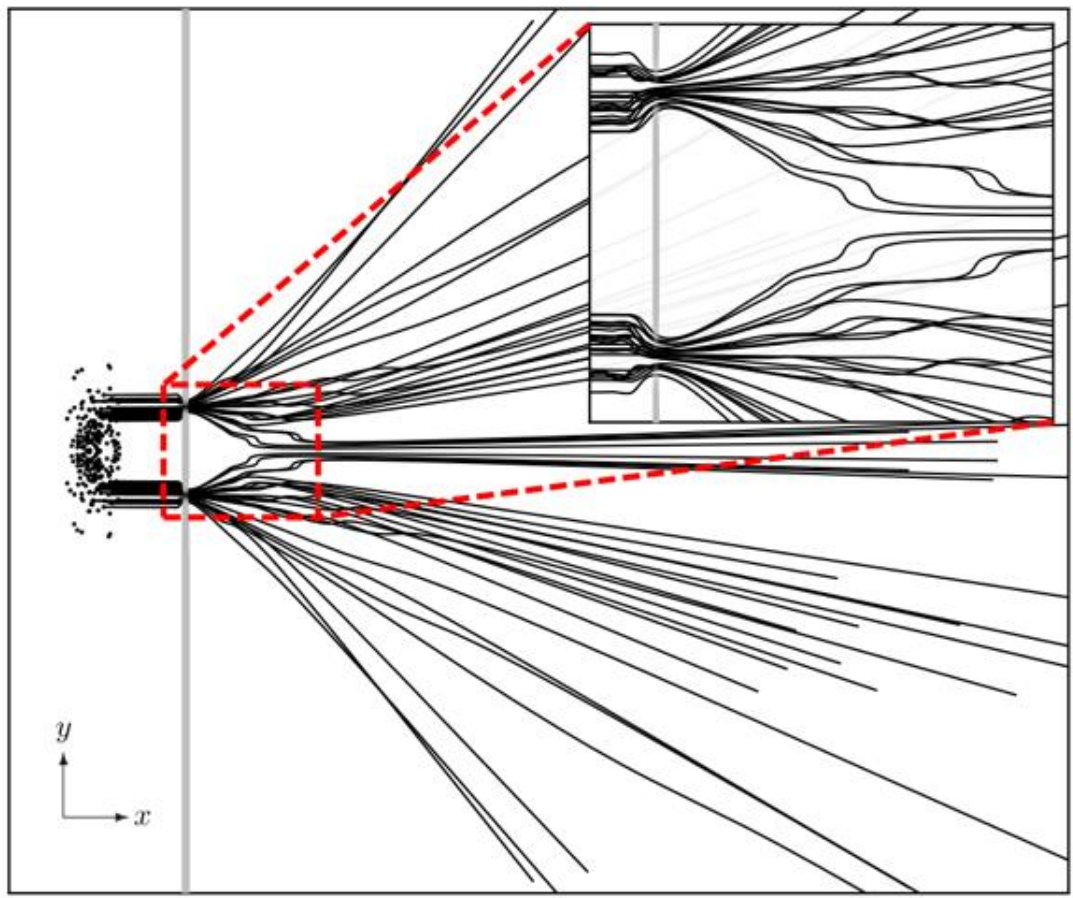

---

[19] Diagram from Das, S., Noth, M. and Dürr, D. (2019) "Exotic Bohmian Arrival Times of Spin ½ Particles" *Phys. Rev. A* 99, 052124, DOI: https://doi.org/10.1103/PhysRevA.99.052124

Figure 1: Pilot Wave Trajectories Without a Recording Screen

Using the squared-amplitude of the initial wavefunction as a probability measure, one gets a probability distribution over the space of trajectories from which one can calculate a probability distribution of *first arrival places and times* of the particle at the geometrical plane of the screen. The screen itself as a physical entity is completely absent from this calculation.

How can one get from this starting point to a prediction for the statistical distribution of where (and when) spots form *on a screen* if the experiment is repeated many times? Here we face a version of Einstein's problem. We are not in a position to provide any exact quantum treatment of the screen and the rest of the apparatus. And even if we could write the relevant equations down, we could not solve them. So it appears that even the pilot wave theory, despite the postulation of particles with trajectories, cannot cleanly make empirical predictions for this experiment from its foundational principles.

Standard operating procedure is to *completely overlook this entire problem*. On first seeing Figure 1 everyone has the immediate reaction that it *provides both a prediction and explanation, from within the pilot wave account, of the observed interference bands in actual two-slit experiments*. The thought underlying this intuitive conclusion is obvious: one supposes that the spots form on the screen *nearly* where and when the trajectories calculated in Figure 1 intercept the geometrical plane where the screen will be put. If that is correct, then we can get from the calculation of the trajectories to the actual data, within some error bounds indicated by "*nearly*".

Let's formulate the reasoning. Everyone agrees that the physical presence of the screen will make *some* difference to the wavefunction of the particle, and hence make some difference to the trajectory of the particle. The actual particle trajectories will deviate from those depicted in Figure 1, especially in the region near the screen. Without a physical analysis, the magnitude of this deviation for a given screen geometry cannot be known. Let's represent it by a factor $\varepsilon_s$ (the subscript indicating "space"). $\varepsilon_s$ represents the average spatial distance from the location where a trajectory depicted in Figure 1 first arrives at the geometrical plane of the screen to the spatial location where a particle originating from the same initial point comes in contact with the actual screen when it is in place. For some screen geometries $\varepsilon_s$ would be massive—for example, a screen that covered one slit but not the other—but no one would use such a screen when looking for interference effects between the slits.

Our second issue concerns the mechanism by which spots form on the screen. Our assumption is that the spots form somewhere close to where the particle first comes near the screen. The mechanism is not likely to be perfectly exact in that respect, so there will be a second factor $\delta_s$ that quantifies the average distance from the point of first contact with the screen to the place where the spot forms. Without a detailed physical analysis of the screen, $\delta_s$ cannot be determined from the theory. And we do not have—and never have had—any such detailed analysis.

Nonetheless, there is an obvious way to proceed. For a given screen, there should be some precise $\varepsilon_s$ and $\delta_s$. So we can say that the observed spatial distribution of spots on the screen should be within $\varepsilon_s + \delta_s$ of the calculated distribution without the screen. Therefore, if $\varepsilon_s + \delta_s$ is sufficiently small we can take the calculated distribution without the screen to be a good first approximation to what we will see with the screen. But if there is no *purely theoretical* justification for believing that $\varepsilon_s + \delta_s$ is small, can there be any justification at all?

At least two sorts of *empirical observations* could be brought forward as evidence that $\varepsilon_s + \delta_s$ is small. Since both $\varepsilon_s$ and $\delta_s$ depend on the precise physical structure of the screen, one could run the experiment repeatedly with differently constructed screens, all of which are regarded as "making position measurements". Either the data from these different runs will be quite similar or notably different. If they are different, then the exact composition of the screen is making an important contribution to the collected data, which is an issue that needs to be addressed. That would mean that the sorts of calculations that are commonly done—which ignore the screen and just use some Hermitian operator or POVM to extract the predictions[20]—

---

[20] As Siddhant Das remarks (p.c.), this is granting too much to the "standard approach" because even without the screen there is no obvious POVM or Hermitian operator to employ. In a *stationary* solution, one could square the wavefunction (in position basis) in the vicinity of the screen and then renormalize to get a probability measure at the

*cannot* be generally reliable. That would not be trouble specifically for the pilot-wave approach, but would taint *all* currently existing treatments.

Alternatively, if differently constructed apparatuses all yield essentially the *same* data, that is reason to believe that $\varepsilon_s + \delta_s$ is small for *all* of them. The data plausibly agree because in all of them the observed data is close to the predicted distribution without the screen in place. So even without any detailed physical analysis of the screens we can have evidence that the recording apparatus is not making a significant contribution to the distribution.

A second, less direct, empirical line of argument can support the supposition that $\varepsilon_s + \delta_s$ is small, namely if *the observed spatial distributions closely match the theoretically derived distribution without the screen.* If those theoretical distributions are close to being accurate, that is evidence for the joint hypotheses that the theory is giving trajectories that are close to correct *and* that the presence of the screen is making only minor changes to the trajectory *and* that the spots are forming quite close to where the particle meets the screen. As with all such cases, it is possible that all the hypotheses are wrong in just the right way to compensate for each other, but *a priori* we rightly regard such a possibility as far-fetched.

Similar remarks could be made about Einstein's problem with the clocks. He did not—and could not—provide a *theoretical* account of how actual clocks work. He could not prove, or even provide plausible grounds to believe, that clocks register (nearly) the proper time along their trajectory. But he could rightfully point out, first of all, that what are regarded as well-made clocks *all agree with each other* (within small bounds) even though they are constructed in quite different ways. That is evidence that they are all measuring *something* objective that is independent of the structure and existence of the clocks. In Relativity, the only candidate for such an objective quantity is the proper time along the trajectory. Then when you add that the predictions Einstein made—using the assumption that physical clocks (approximately) measure

---

screen, but in the actual experiment done "one particle at a time" the wavefunction is certainly not stationary and there is no obvious way to accumulate the probabilities pegged to specific times over the entire period of data collection to calculate a net chance of a spot forming at any specific location. For example, if one tries to model the situation as a rapid series of distinct instantaneous "position measurements" taken at the screen, one runs into the quantum Zeno paradox and the prediction is that no spot will ever form.

proper time—turn out to be so accurate, you get strong holistic confirmation for *both* the fundamental theory *and* the assumption about clocks. It is not a proof, but it is about as good as it gets in empirical science.

Contrast this situation with the "standard approach" to quantum theory where there are no particles, no trajectories, and no physical analysis of the recording equipment. There are rules-of-thumb about how to squeeze statistical predictions out of the theory, unbacked by any theory. A tremendous amount of work is done by the association of physical apparatuses with Hermitian operators or POVMs without any general account of how or why such an association should be made. There is not even a vague sketch of how physical spots on screens form.

All of these observations about "standard quantum theory" are so common as to be trite, but the analysis points toward a possible way to—in Kuhn's terminology—promote a puzzle into a crisis. Note: in the pilot-wave theory using the $\varepsilon_s + \delta_s$ approach, *exactly the same type of analysis can be offered for arrival time data* (*when* the spot forms on the screen) *as for locational data of where the spot forms on the screen*. The physical presence of the screen might delay or accelerate the arrival of the particle, yielding a factor $\varepsilon_t$. And the mechanism by which a clock records a time can create a deviation from the time that the particle did arrive, yielding a factor $\delta_t$. But if $\varepsilon_t + \delta_t$ is small (which can be subjected to empirical test using different clocks in exactly the same way), then the clean calculations of arrival *times* without the screen in place can form the basis for predictions of arrival time distributions, and indeed of arrival-time-and-place *joint* distributions, which can be experimentally tested.

In contrast, when it comes to arrival times, the standard theory is dead in the water. For—as is well known—there is no "time operator" to play an analogous role in the prediction of arrival time distributions as the "position operator" plays in the prediction of the distribution of locations[21], much less a joint-time-and-position operator. But we can—with present

---

[21] There are technical difficulties even for predicting the statistics of the locations of spots in the standard approach. That approach tells us to express the wavefunction of the "particle" in position basis and then square the amplitudes to get the probabilities of the "location" that will be indicated. But that operation is carried out only on the wavefunction *at a single moment*. Since we don't know when the spot will form, we equally don't know which moment to carry out the operation, nor how to update the wavefunction if no spot forms on the screen at some

technology—acquire joint distributions of locations and times of spots forming on screens, which we have no way *at all* to predict in the standard setting.[22] In contrast, the pilot-wave approach together with the assumption that $\varepsilon_s + \delta_s$ and $\varepsilon_t + \delta_t$ are both small does make predictions. And other approaches concerning how to deal with our two problems may produce other predictions from the pilot-wave foundation.[23]

It is not in the remit of this essay to solve the problems of how to extract predictions from the quantum formalism in these circumstances, or to survey all the proposals in the literature for how to make these predictions.[24] It is rather to insist that these problems exist, that they concern the outcomes of table-top experiments that are within reach of presently existing technology, and that the problems arise from a systematic methodological difficulty that has always infected actual physics, including Relativity. Nonetheless, they are not widely appreciated.

There may be many reasons for this, but plausibly one is the mistaken idea that flesh-and-blood physicists are just somewhat flawed versions of Bošković's demon. The demon is supposed to be able to make predictions from only the fundamental physical principles, without any need for additional questionable stipulations or uncashed promissory notes. One might grant that actual physicists do use idealizations and approximations, but only as a means to derive (approximate) answers for the very same calculations that the demon does effortlessly. But as Einstein's remarks illustrate, the situation for actual scientists is much further from the position

---

moment. This is connected with the fact that screens are *passive detectors*, in Will Cavendish's terminology. The instrumentation does not determine when the outcome will occur. One wants to say that they particle itself does, but on the standard approach there is no particle to play that role.

[22] See Fig. 1 and 6 of Das. S, Deckert, D.-A., Kellers, L., Krekels, S. and Struyve, W. ibid, for experimental data for such joint distributions.

[23] Concerning the effect of the detecting apparatus on the target system wavefunction, one proposal is to represent that by a boundary condition, as the barrier with slits is represented by boundary conditions. Instead of a steep potential, which will reflect the wavefunction, one uses a completely Absorbing Boundary Condition (ABC). But if implemented exactly it would have some striking *empirical* consequences, which seem to be false. See Cavendish, W. and Das, S. (2025) "Absorbing detectors meet scattering theory", https://arxiv.org/abs/2509.07518 to appear in *Phys. Rev. A*: https://doi.org/10.1103/332p-4nl1.

[24] Will Cavendish has surveyed much of the presently existing literature on this topic in ``An unexpected shared prediction of proposed solutions to the arrival time problem". See also Muga, J. G. and Leavens, C. R. (2000) "Arrival time in quantum mechanics" *Physics Reports* 338, Issue 4, 253-438 and Siddhant Das's PhD thesis "Arrival time distributions and spin in quantum mechanics" https://pure.mpg.de/pubman/faces/ViewItemOverviewPage.jsp?itemId=item_3652179.

of the demon than that suggests. The stipulations serve an essential role in making *any predictions at all*, and the content of the stipulations is not determined by any derivation—even an approximate one—from known fundamental principles. Without appreciating this we will not understand how "quantum theory", or any actual physical theory in history, has made contact with experimental data and hence how empirical testing of those theories actually takes place. In the specific case of quantum theory, there are compelling reasons to think that the shape that this problem takes for different "interpretations" can be radically different, as we have seen. In comparing the various "interpretations" the exact nature, theoretical justification, and *empirical justification* of these essential parts of the predictive apparatus should be a central topic of investigation and analysis.[25]

---

[25] Thanks to Siddhant Das for comments on this manuscript.